\documentclass[aps,prl,preprint,groupedaddress]{revtex4}
\def\lap{\hbox{${_{\displaystyle<}\atop^{\displaystyle\sim}}$}}
\def\gap{\hbox{${_{\displaystyle>}\atop^{\displaystyle\sim}}$}}

\def\lap{\hbox{${_{\displaystyle<}\atop^{\displaystyle\sim}}$}}
\def\gap{\hbox{${_{\displaystyle>}\atop^{\displaystyle\sim}}$}}

\newcommand{\nd}        {\mbox{\boldmath$n_{d}$}}

\newcommand{\bfI}       {\mbox{\boldmath $I$}}
\newcommand{\bfdelta}   {\mbox{\boldmath $\delta$}}
\newcommand{\Om}        {\mbox{\boldmath $\Omega$}}
\newcommand{\nom}       {\mbox{\boldmath$n_{\Omega}$}}

\newcommand{\fbf}         {\mbox{\boldmath$f$}}
\newcommand{\rbf}         {\mbox{\boldmath$r$}}

\newcommand{\Ln}         {\mbox{\boldmath$L_n$}}
\newcommand{\Lc}         {\mbox{\boldmath$L_c$}}
\newcommand{\Ltot}         {\mbox{\boldmath$L$}}
\newcommand{\kappavec} {\mbox{\boldmath$\kappa$}}
\newcommand{\Omn} {\mbox{\boldmath$\Omega_n$}}
\newcommand{\Omc} {\mbox{\boldmath$\Omega_c$}}
\newcommand{\lambdastar} {\mbox{\boldmath$\lambda^*$}}

\usepackage{graphicx}

\begin{document}

%\preprint{}

\title{Constraining Hadronic Superfluidity with Neutron Star Precession} 


\author{Bennett Link}
\affiliation{Montana State University, Department of Physics, Bozeman MT
59717}
\email[]{blink@dante.physics.montana.edu}
%\thanks{}


\date{\today}

\begin{abstract}
I show that the standard picture of the neutron star core containing
coexisting neutron and proton superfluids, with the proton component
forming a type II superconductor threaded by flux tubes, is
inconsistent with observations of long-period ($\sim 1$ yr) precession
in isolated pulsars. I conclude that either the two superfluids
coexist nowhere in the stellar core, or the core is a type I
superconductor rather than type II. Either possibility would have
interesting implications for neutron star cooling and theories of spin
jumps (glitches).
\bigskip\bigskip
\centerline{\it Physical Review Letters, in press}
\end{abstract}

%\keywords{}

\bigskip

\maketitle

%\section{}
%\subsection{}
%\subsubsection{}

Neutron superfluidity and proton superconductivity in neutron stars
could have a number of interesting consequences for observed spin
behavior and thermal evolution. Interaction of superfluid vorticity
with the nuclei of the inner crust or superconducting flux tubes in
the core could lead to the jumps in spin rate, {\em glitches}, seen in
many neutron stars \cite{lss}. The specific heat of the stellar
interior is determined by the state of the matter, while neutrino
emission processes which cool a young neutron star are strongly
suppressed in the presence of hadronic superfluids (see, e.g.,
\cite{tsuruta}). The properties of condensed hadronic systems are also
of interest in studies of heavy nuclei near the neutron drip line
\cite{heavy} and light halo nuclei \cite{halo}. The properties of
hadronic systems in beta equilibrium is therefore a central problem in
both nuclear astrophysics and nuclear physics.

Reliable predictions of the pairing states of the neutron star core
are not yet possible as they require extrapolation of nucleon-nucleon
potentials well above nuclear saturation density, $\rho_s\equiv
2.8\times 10^{14}$ g cm$^{-3}$. The current picture of the neutron
star interior posits that the outer core consists of mostly $^3P_2$ or
$^3P_2$-$^3F_2$ superfluid neutrons, with about 5\% of the mass in
type II $^1S_0$ superconducting protons, normal electrons and fewer
muons
\cite{hoffberg,chao,amundsen,wambach,takatsuka,elgaroytriplet,moderngap}.
Above a density $\simeq 1.7\rho_s$, the pairing situation is
essentially unknown \cite{moderngap}.

The compelling evidence for precession in isolated pulsars
\cite{stairs,cordes93,shabanova} provides new probes of the state of a
neutron star's exotic interior. The periodic timing behavior of PSR
B1828-11 and correlated changes in beam profile have been interpreted
as due to precession with a period of $\sim 1$ yr and an amplitude of
$\simeq 3^\circ$~\cite{linkepstein,ja,rezania1}. The measured
precession period implies a fractional distortion of the star (in
addition to its rotational distortion) of $\epsilon\simeq
10^{-8}$. This deformation could be sustained by magnetic stresses
\cite{wasserman,cutler}, crust stresses \cite{cul}, or a combination
of the two. 

The picture of the outer core I will consider is as follows. The
neutron fluid rotates by establishing a triangular array of quantized
vortex lines, parallel to the axis of the angular momentum of the
superfluid and with an areal density of $n_v =
2m_n\Omega_n/\pi\hbar\simeq 10^4 P({\rm s})^{-1}\mbox{ cm}^{-2}$,
where $m_n$ is the neutron mass, $\Omega_n$ is the angular velocity of
the superfluid and $P$ is the spin period. The average vortex spacing
is $l_v\equiv n_v^{-1/2}\simeq 10^{-2} \hbox{$P$(s) cm}$.  If magnetic
flux penetrates the superconducting core, it is organized into
quantized flux tubes, with an areal density $n_\Phi = B/\Phi_0\sim
10^{19} B_{12} \mbox{ cm}^{-2}$, where $B_{12}\equiv 10^{12}B$, $B$ is
the average core field in Gauss and $\Phi_0\equiv\pi\hbar c/e=2\times
10^{-7} \mbox { G cm}^{-2}$ is the flux quantum. The average spacing
between flux tubes is $l_\Phi\equiv n_\Phi^{-1/2}\simeq 3000
B_{12}^{-1/2}\hbox{ fm}$.  The magnetic field in the core of a flux
tube is approximately the lower critical field for the superconducting
transition, $H_{c1}(\simeq 10^{15}$ G). Unlike the vortex array, which
is expected to be nearly rectilinear, the flux tube array is likely to
have a very complicated and twisted structure \cite{rzc}. Hence, the
vortices are entangled in the far more numerous flux tubes. In this
Letter I show that this entanglement restricts the precession to be of
very high frequency and low amplitude, in conflict with observations.

A flux tube has a core of normal protons; the radius of this region is
of order the proton coherence length $ \xi_p \simeq 30 $ fm. Outside
the flux tube, the magnetic field falls off exponentially over a
distance equal to the London length: $ \Lambda_p \simeq 80 $ fm.
(Type II superconductivity occurs when $\sqrt{2}\Lambda_p>\xi_p$).  A
vortex has a core of normal neutrons, of characteristic radius 
the neutron coherence length, $\xi_n\sim 10$ fm. Entrainment of
protons in the neutron flow about a vortex magnetizes the vortex
\cite{als}. The length scale over which the vortex's magnetic field
decays is $ \Lambda_n \simeq 10 $ fm.  These length scales are
for typical parameters of the core: a superconducting transition
temperature of $10^9$ K, an effective proton mass of
half the bare mass, a proton mass fraction of 0.05 and a total mass
density of $3\times 10^{14}$ g cm$^{-3}$~\cite{sjoberg}. The protons
do not rotate by establishing an array of vortices, but corotate with
the crust and electron fluid approximately as a rigid body by
adjusting the London current \cite{als}. 

For the angular momentum of the neutron fluid to change, the vortices
must move. The flux tubes in which they are entangled, however, impede
their motion. As a vortex segment approaches a flux tube segment, the
total magnetic energy increases (decreases) if the vortex and flux
tube are aligned (anti-aligned). The energy per intersection is
$E_{\rm int}\sim 5$ MeV, with a range $\sim\Lambda_p$
\cite{ccd,rzc}. The vortices are effectively pinned against the flux
array, unless they can push the flux tubes through the star, or cut
through them by surmounting the numerous energy barriers. The pinning
force is $F_p\equiv E_{\rm int}/\Lambda_p\sim 0.1$ MeV fm$^{-1}$ per
intersection, corresponding to a force per unit length of vortex of $
f_p \sim F_pn_\Phi^{1/2}= 3\times 10^{15} B_{12}^{1/2}$ dyn cm$^{-1}$.
Because the core is highly conductive, vortices can push the flux
tubes only very slowly \cite{rzc}. Hence, unless the vortices cut through
the flux tubes, the core flux tubes and vortices move together, and
the crust, which is frozen to the strong field emerging from the core,
approximately follows the motion of the charged fluid. 

The charged fluid of the core responds to changes in its rotation rate
approximately as a rigid body if magnetic stresses are sufficient to
enforce corotation. In the superconducting core, magnetic stresses
propagate as cyclotron-vortex waves of frequency $ \omega_{cv}=
(H_{c1}B/4\pi\rho_p)^{1/2} k$, where $k$ is the excitation wavenumber
\cite{mendell}. Taking $k= \pi/R$, where $R$ is the stellar radius,
$\rho_p = 1.5\times 10^{13}$ g cm$^{-3}$ and $B=5\times 10^{12}$ G,
gives the characteristic frequency at which magnetic stresses are
communicated through the core, $\omega_{cv,0}\simeq 10$ rad
s$^{-1}$. The frequency $\omega_{cv,0}$ represents an approximate
upper limit to the precession frequency of the star as a whole.

To calculate the precession dynamics, I assume that pressure gradients
and magnetic stresses force the crust and the charged fluid to move
together, and refer to the charged core fluid plus crust, whose spin
rate we observe, as the ``body'', even though it contains $\lap 10$\%
of the star's mass. By the arguments given above, the core neutron
fluid, which accounts for $\gap 90\%$ of the stellar mass, has its
angular momentum fixed to this body, through pinning of the vortices
to the flux tubes. I assume that the $^1S_0$ vortices of the inner
crust are not pinned to nuclei, and nearly follow the body's rotation
axis so that they have a negligible effect on the precession dynamics
\cite{lc}. With these idealizations, let us write the the inertia
tensor of the charged fluid as the sum of a spherical piece, a
centrifugal bulge that follows the instantaneous angular velocity and
an oblate, biaxial deformation bulge aligned with the body's principal
axis: \begin{equation} \bfI_c = I_{0,c} \bfdelta + \Delta I_\Omega
\left (\nom\nom - {1\over 3}\bfdelta\right ) + \Delta I_d \left
(\nd\nd- {1\over 3}\bfdelta\right ).  \end{equation} Here $I_{0,c}$ is
the moment of inertia of the charged fluid (plus any components
tightly coupled to it) when non-rotating and spherical, $\bfdelta$ is
the unit tensor, $\nom$ is a unit vector along the body's angular
velocity $\Om_c$, $\nd$ is a unit vector along the principal axis of
inertia, $\Delta I_\Omega$ is the increase in oblateness about $\Om_c$
due to rotation and $\Delta I_d$ is the portion of the body's
deformation that is frozen in the body. Let the neutron fluid's
angular momentum vector $\Ln$ be perfectly tied to the core flux tube
array, so that $\Ln$ is fixed with respect to the body. The total
angular momentum is $\Ltot=\Ln+\Lc$, where $\Lc$ is the total angular
momentum of the body. The Euler equations in the body frame are
$\bfI_c\cdot\dot{\Om}_c + \Omc\times(\Lc + \Ln) = 0.$ Define principal
axes in the body ($x_1$, $x_2$, $x_3$), where $x_3$ is along the major
principal axis ($\hat{x}_3=\nd$). The principal moments of inertia are
$I_1 = I_0 + 2\Delta I_\Omega/3 - \Delta I_d/3 = I_2 =
I_3(1+\epsilon)^{-1}$, where $\epsilon\equiv\Delta I_d/I_1>0$. Let the
angle between $x_3$ and $\Ln$ be $\alpha$. If $\Lc$, $\Ln$ and $\Omc$
are all aligned, the star is in a state of minimum energy for a given
angular momentum and does not precess. A likely precessional state is
one in which $\Lc$ and $\Ln$ are perturbed slightly about this
stationary point. To define angles, let $x_3$, $\Ln$, $\Lc$ and $\Omc$
all lie in a plane at $t=0$, with $\theta$ the angle between $x_3$ and
$\Ltot$, and $\theta^\prime$ the angle between $\Omc$ and $\Ltot$ (see
Fig. 1).  Linearizing Euler's equations in $\Omega_{c1}$,
$\Omega_{c2}$ and $\alpha$, gives the solutions: 
\begin{equation}
\Omega_{c1}(t) = A\cos\Omega_pt - \theta_0\Omega_c , \qquad
\Omega_{c2}(t) = A\sin\Omega_pt, 
\end{equation} 
where $\Omega_p \equiv
\epsilon\Omega_c + L_n/I_1$ is the body-frame precession frequency,
and, \begin{equation} {A\over\Omega_c} = \alpha {L_n\over I_1}\left
[{1\over\Omega_c} + {1\over\Omega_p}\right ] - \theta\left
[1+{\Omega_p\over\Omega_c}\right ], \quad \theta_0 = \alpha {L_n\over
I_1\Omega_p}.  \end{equation} The motion of $\Omc$ is a circle of
angular radius $\vert A\vert/\Omega_c$, about an axis that takes an
angle $\theta_0$ with respect to $x_3$ in the $x_1-x_3$ plane,
completing one revolution in a time $2\pi/\Omega_p$. Removing the
pinned component ($L_n=0$) gives the familiar result of
$\Omega_p=\epsilon\Omega_c$, $A/\Omega_c\simeq\theta$.  Restoring the
pinned neutron fluid, and taking $L_n\simeq I_n\Omega_c$ gives
$\Omega_p\simeq (I_n/I_1)\Omega_c\simeq 10\Omega_c$, independent of
$\alpha$ and $\theta$. When $\Omega_p$ exceeds $\omega_{cv,0}$, as for
PSR B1828-11, the precession frequency is likely to be closer to
$\omega_{cv,0}\simeq 10$ rad s$^{-1}$, still very high. {\em As long
as $\Ln$ is pinned to the body, the star precesses at high frequency
for any finite $\alpha$ or $\theta$}.

First consider a state in which $\Ln$ and $\Lc$ are both aligned at $t=0$, but
$\Omc$ is not along $\Ltot$. In this case, $\alpha=\theta$, giving
\begin{equation}
{A\over\Omega_c}  =  - \epsilon\theta\left
(1+{\Omega_c\over\Omega_p}\right ),
\qquad
\theta_0  =  \theta\left (1-\epsilon{\Omega_c\over\Omega_p}\right ). 
\end{equation}
The angular velocity vector of the body takes a tiny circle of
(angular) radius $\simeq \epsilon\theta<<\theta$ about axis $o$ in
Fig. 1, nearly coincident with $\Ltot$. Since $\Ln$ is fixed in the
body, it too goes around $o$. For the wobble angle to be
$\sim 3^\circ$ with $\Ln$ fixed in the body, $\alpha<<\theta$ is
obviously required. Suppose, for example, that $\alpha=0$, that is,
$\Ln$ is aligned with $x_3$. Then
\begin{equation}
{A\over\Omega_c}  =  - \theta\left
(1+{\Omega_p\over\Omega_c}\right ) =
-(\theta+\theta^\prime) \equiv -\beta,
\qquad
\theta_0  =  0.
\end{equation}
Now the angular velocity of the body takes a circle about $x_3$, with
radius $\beta\sim \theta$. Relatively large-amplitude precession
occurs, but still at very high frequency.  For precession of large
amplitude to occur at low frequency, $\Ln$ must be able to closely
follow the rotation axis of the body, so that the $\Omc\times\Ln$ term
in Euler's equations becomes small compared to $\Omc\times\Lc$. For
this to happen, the neutron vortices must be able cut through the flux
tubes. For $\beta\simeq 1^\circ$, this is likely. The flow of the
neutron superfluid past a vortex pinned against a flux tube creates a
Magnus force per unit length of vortex at location $\rbf$ of $\fbf_m =
\rho_n\kappavec\times ([\Omn-\Omc]\times \rbf )$, where $\rho_n$ is
the mass density of the neutron superfluid and $\kappavec$ is a vector
in the direction of the neutron vorticity with magnitude $h/2m_n$
\cite{shaham77}.  For simplicity, take $\kappavec=\kappa\hat{x}_3$ and
$\Omn=\Omega_n\hat{x}_3$ and $\Omega_n=\Omega_c\equiv\Omega$. At
$t=0$, when $\Omn$, $\Omc$ and $\Ltot$ all lie in the $x_1-x_3$ plane,
the angular velocity of the body is $\Omc=\Omega(-\sin\beta\,\hat{x}_1
+ \cos\beta\, \hat{x}_3)$. The instantaneous Magnus force per unit
length of vortex as a function of position in the star is, for small
angles, $\fbf_m = -\hat{x}_1\rho_n\kappa\Omega\beta x_3$.
If $f_m$ exceeds $f_p$, the pinning force per unit length on a typical
vortex, the vortices will cut through the flux tubes
that are in their way. This condition gives 
$\vert x_3\vert > f_p/\rho_n\kappa\Omega\beta$. 
For $\Omega=16$ rad s$^{-1}$ (PSR B1828-11), the inferred $\beta$ of
3$^\circ$ and a density $\rho_s=3\times 10^{14}$ g cm$^{-3}$; 
$\vert\fbf_m\vert$ exceeds $f_p=10^{16}$ dyne cm$^{-1}$ for $\vert
x_3\vert > 2\times 10^{-2}R$, that is, the Magnus force will force the
vortices through the flux tubes almost everywhere in the star. This
process is highly dissipative.

As a vortex is forced through a flux tube, quantized vortex waves,
{\em kelvons}, are excited, which propagate along the vortex and
eventually dissipate as heat \cite{eb}. In the rest frame
of a straight vortex along the $\hat{z}$ axis, suppose a straight flux
tube in the $y-z$ plane approaches at speed $v$. Since
$\Lambda_n<\Lambda_p$, the finite (magnetic) radius of the vortex can be
ignored.  Let the vortex and flux tube overlap at $t=0$. The vector
separation between a point at the center of the flux tube which will
coincide with the vortex at $t=0$ is ${\mathbf s}(t)=vt\hat{x}$. As a
simple model of the interaction force, consider 
${\mathbf f_{\rm int}}({\mathbf s}(t)) = F_p(s/\Lambda_p){\rm
  exp}[(1-s^2/\Lambda_p^2)/2]\delta (z) \hat{x}$, 
where $\delta(z)$, the Dirac-delta function, gives the distribution of
the interaction force along the vortex (justified below). 

The relative velocity between vortices and flux tubes is $v\simeq
R\Omega_c\beta$ in the initial stage that flux tubes cut through
vortices; taking $\beta$ comparable to the observed wobble angle gives
$v=10^6$ cm s$^{-1}$ for PSR B1828-11. As a flux tube passes through a
vortex, it excites kelvons of characteristic frequency
$\omega_0=v/\Lambda_p$.  Kelvons on a free vortex are circularly
polarized waves. The frequency of a kelvon is related to its
wavenumber by $ \omega_k = \hbar k^2/2\mu $ where $\mu$ is the
effective mass of a kelvon, given by $\mu = m_n/\pi\Lambda$.  The
dimensionless parameter $\Lambda$ is $\simeq 0.116 -\ln (k\xi_n)$ for
wavenumbers in the range $l_v^{-1}<<k<<\xi_n^{-1}$ \cite{sonin}, which
is easily satisfied for the characteristic wavenumber
$k_0=(2\mu\omega_0/\hbar)^{1/2}$ of interest. For $v=10^6$ cm s$^{-1}$
and $\xi_n=10$ fm, these relationships give $k_0=5\times 10^{-4}$ fm
and $\mu=0.06 m_n$.

The total energy transferred to
a vortex per scattering in a potential is given in first-order
perturbation theory by \cite{eb}
\begin{equation}
\Delta E = {\hbar\over 4\pi\rho_n\kappa\mu}\int_{-\infty}^\infty dk\,
k^2\, 
\left |\int_{-\infty}^\infty dz\, {\rm e}^{-ikz} 
\int_{-\infty}^\infty dt\, {\rm e}^{i\omega_k t} f_+(z,t)\right |^2. 
\label{dE}
\end{equation}
Here $ f_+(z,t) \equiv {\mathbf f_{\rm int}}(z,t)\cdot\lambdastar $
where $\lambdastar\equiv (\hat{x}+i\hat{y})/\sqrt{2}$ is the
right-circularly polarized unit vector for a kelvon.  Since
$k_0\Lambda_p<<1$, the flux tube exerts a highly localized force along
the length of the vortex, justifying the use of the $\delta$-function
in model for the force. To estimate the total
dissipation rate in the core, take $l_\Phi\equiv
n_\Phi^{-1/2}=(\Phi_0/B)^{1/2}$ as the average distance between
intersections of a vortex line with a flux tube. The total number of
vortices in the core is $N=2\pi R^2\Omega_n/\kappa$. 
Taking a typical vortex length of $R$, gives a total
dissipation rate in the core of $\dot{E} \gap  N n_\Phi R v\Delta
E$, or 
\begin{equation}
{dE\over dt}\,
\gap 4{F_p^2 R^3\Omega_n B\over
\rho_n\kappa^2\Phi_0} \left ({2\mu\Lambda_p v\over\hbar}\right )^{1/2}, 
\label{edot}
\end{equation}
a lower limit, since the excitation of flux tubes, which is also
dissipative, was ignored.  Different choices for the dependence of the
interaction force on ${\mathbf s}$ give the same scaling on the
parameters appearing in eq. (\ref{edot}), with slightly different
numerical factors.  The use of eq. (\ref{dE}) assumes that cuttings of
the vortex at different locations can be treated as separate events,
with the excitations due to different cuttings adding
incoherently. This will be the case as long as $k_0l_\Phi\gg 1$. For
$v=10^6$ cm s$^{-1}$ and $B_{12}=1$, $k_0l_\Phi$ is $\sim 2$, so the
approximation of kelvons as distinct wavepackets is a somewhat crude
one, but the lower limit in eq. (\ref{edot}) should be a reasonable
estimate for the velocities of interest.  Taking $v=10^6$ cm s$^{-1}$,
$\mu=0.06m_n$, $R=10$ km $\Lambda_p=80$ fm and $F_p=0.1$ MeV fm$^{-1}$
gives a dissipation rate of $dE/dt\sim 10^{41}$ erg s$^{-1}$. Now
consider the excess rotational
energy of the precessing star. The energy in the body $E_{\rm rot}$
is related to the energy in the inertial frame $E_0$ by $ E_{\rm rot}
= E_0 - \Ltot\cdot\Om$. Most of the angular momentum is in the
neutrons, so $\Ltot\simeq\Ln$. The excess rotational energy is thus
$\Delta E_{\rm rot} \simeq I_n\Omega_n^2\beta^2/2 \simeq 2\times
10^{44}$ erg. The characteristic damping time is $ \tau_d\equiv \Delta
E_{\rm rot}(dE/dt)^{-1}\lap 1 $ hr. Over this short timescale, the
precession damps to small amplitude. When $\beta$ is $\lap 0.06^\circ$, the
Magnus force cannot drive the vortices though the flux tubes anywhere
in the star; $\Ln$ is now fixed in the body, and therefore cannot
follow the total angular momentum, so the star precesses at frequency
$\Omega_p\simeq\omega_{cv,0}\simeq 10$ rad s$^{-1}$. In general, then,
{\em long-period precession is not possible.}

To summarize, these estimates show that a neutron star core containing
coexisting neutron vortices and proton flux tubes cannot precess with
a period of $\sim 1$ yr.  Since $\Omega_p=\epsilon\Omega_c+L_n/I_1$,
the fraction of the neutron component's moment of inertia that is
pinned against flux tubes must be $\ll\epsilon\simeq 10^{-8}$.  Hence,
observations require that {\em neutron vortices and proton flux tubes
coexist nowhere in the star}. Either the star's magnetic field does
not penetrate any part of the core that is a type II superconductor,
which seems highly unlikely, or at least one of the hadronic fluids is
not superfluid. This latter possibility appears unlikely
in the face of pairing calculations which predict coexisting neutron
and proton superfluids in the outer core
\cite{hoffberg,chao,amundsen,wambach,takatsuka,elgaroytriplet,moderngap}. 

If the core is a type I superconductor, at least in those regions
containing vortices, the magnetic flux could exist in macroscopic
normal regions that surround superconducting regions that carry no
flux. In this case, the magnetic field would not represent the
impediment to the motion of vortices that flux tubes do, and the star
could precess with a long period. Perhaps PSR B1828-11 and other
precession candidates are giving us the first clue that neutron stars
contain a type I superconductor. Another, strange possibility, is that
``neutron stars'' are in fact composed of strange quark matter
\cite{afo}.

The possibilities discussed above have interesting implications for
models of neutron star spin and thermal evolution.  Glitch models that
rely on vortex-flux tube interactions, e.g., \cite{rzc}, would no
longer apply, leaving the inner crust superfluid as a possible origin
of glitches \endnote{Evidence that glitches originate in the inner crust
follows from angular momentum considerations showing that glitches
involve $\sim 1\%$ of the star's moment of inertia on average in most
neutron stars \cite{lel}.}. The URCA reactions, which are strongly
suppressed in regions where both neutrons and protons are superfluid,
could be significantly increased if macroscopic regions of the core
are normal, affecting the thermal evolution of young neutron stars.

\begin{acknowledgments}
I thank R. I. Epstein, C. Thompson and I. Wasserman for valuable discussions. 
This work was supported by the National Science Foundation under
Grant No. AST-0098728. 
\end{acknowledgments}


\bibliography{references}

\begin{thebibliography}{31}
\expandafter\ifx\csname natexlab\endcsname\relax\def\natexlab#1{#1}\fi
\expandafter\ifx\csname bibnamefont\endcsname\relax
  \def\bibnamefont#1{#1}\fi
\expandafter\ifx\csname bibfnamefont\endcsname\relax
  \def\bibfnamefont#1{#1}\fi
\expandafter\ifx\csname citenamefont\endcsname\relax
  \def\citenamefont#1{#1}\fi
\expandafter\ifx\csname url\endcsname\relax
  \def\url#1{\texttt{#1}}\fi
\expandafter\ifx\csname urlprefix\endcsname\relax\def\urlprefix{URL }\fi
\providecommand{\bibinfo}[2]{#2}
\providecommand{\eprint}[2][]{\url{#2}}

\bibitem[{\citenamefont{Lyne et~al.}(2000)\citenamefont{Lyne, Shemar, and
  Smith}}]{lss}
\bibinfo{author}{\bibfnamefont{A.~G.} \bibnamefont{Lyne}},
  \bibinfo{author}{\bibfnamefont{S.~L.} \bibnamefont{Shemar}},
  \bibnamefont{and} \bibinfo{author}{\bibfnamefont{F.~G.} \bibnamefont{Smith}},
  \bibinfo{journal}{Mon. Not. Roy. Astr. Soc.} \textbf{\bibinfo{volume}{315}},
  \bibinfo{pages}{534} (\bibinfo{year}{2000}).

\bibitem[{\citenamefont{Tsuruta}(1998)}]{tsuruta}
\bibinfo{author}{\bibfnamefont{S.}~\bibnamefont{Tsuruta}},
  \bibinfo{journal}{Phys. Rep.} \textbf{\bibinfo{volume}{292}},
  \bibinfo{pages}{1} (\bibinfo{year}{1998}).

\bibitem[{\citenamefont{M{\"u}ller and Sherril}(1993)}]{heavy}
\bibinfo{author}{\bibfnamefont{A.~C.} \bibnamefont{M{\"u}ller}}
  \bibnamefont{and} \bibinfo{author}{\bibfnamefont{B.~M.}
  \bibnamefont{Sherril}}, \bibinfo{journal}{Ann. Rev. Nucl. Part. Phys.}
  \textbf{\bibinfo{volume}{43}}, \bibinfo{pages}{529} (\bibinfo{year}{1993}).

\bibitem[{\citenamefont{Riisager}(1994)}]{halo}
\bibinfo{author}{\bibfnamefont{K.}~\bibnamefont{Riisager}},
  \bibinfo{journal}{Rev. Mod. Phys.} \textbf{\bibinfo{volume}{66}},
  \bibinfo{pages}{1105} (\bibinfo{year}{1994}).

\bibitem[{\citenamefont{Hoffberg et~al.}(1970)\citenamefont{Hoffberg,
  Glassgold, Richardson, and Ruderman}}]{hoffberg}
\bibinfo{author}{\bibfnamefont{M.}~\bibnamefont{Hoffberg}},
  \bibinfo{author}{\bibfnamefont{A.~E.} \bibnamefont{Glassgold}},
  \bibinfo{author}{\bibfnamefont{R.~W.} \bibnamefont{Richardson}},
  \bibnamefont{and} \bibinfo{author}{\bibfnamefont{M.}~\bibnamefont{Ruderman}},
  \bibinfo{journal}{Phys. Rev. Lett.} \textbf{\bibinfo{volume}{24}},
  \bibinfo{pages}{775} (\bibinfo{year}{1970}).

\bibitem[{\citenamefont{Chao et~al.}(1972)\citenamefont{Chao, Clark, and
  Yang}}]{chao}
\bibinfo{author}{\bibfnamefont{N.~C.} \bibnamefont{Chao}},
  \bibinfo{author}{\bibfnamefont{J.~W.} \bibnamefont{Clark}}, \bibnamefont{and}
  \bibinfo{author}{\bibfnamefont{C.~H.} \bibnamefont{Yang}},
  \bibinfo{journal}{Nucl. Phys. A} \textbf{\bibinfo{volume}{179}},
  \bibinfo{pages}{320} (\bibinfo{year}{1972}).

\bibitem[{\citenamefont{Amundsen and {\O}stgaard}(1985)}]{amundsen}
\bibinfo{author}{\bibfnamefont{L.}~\bibnamefont{Amundsen}} \bibnamefont{and}
  \bibinfo{author}{\bibfnamefont{E.}~\bibnamefont{{\O}stgaard}},
  \bibinfo{journal}{Nucl. Phys. A} \textbf{\bibinfo{volume}{437}},
  \bibinfo{pages}{487} (\bibinfo{year}{1985}).

\bibitem[{\citenamefont{Wambach et~al.}(1993)\citenamefont{Wambach, Ainsworth,
  and Pines}}]{wambach}
\bibinfo{author}{\bibfnamefont{J.}~\bibnamefont{Wambach}},
  \bibinfo{author}{\bibfnamefont{T.~L.} \bibnamefont{Ainsworth}},
  \bibnamefont{and} \bibinfo{author}{\bibfnamefont{D.}~\bibnamefont{Pines}},
  \bibinfo{journal}{Nucl. Phys. A} \textbf{\bibinfo{volume}{555}},
  \bibinfo{pages}{128} (\bibinfo{year}{1993}).

\bibitem[{\citenamefont{Takatsuka}(1984)}]{takatsuka}
\bibinfo{author}{\bibfnamefont{T.}~\bibnamefont{Takatsuka}},
  \bibinfo{journal}{Prog. Theor. Phys. Suppl.} \textbf{\bibinfo{volume}{71}},
  \bibinfo{pages}{1432} (\bibinfo{year}{1984}).

\bibitem[{\citenamefont{Elgar{\o}y et~al.}(1996)\citenamefont{Elgar{\o}y,
  Engvik, Hjorth-Jensen, and Osnes}}]{elgaroytriplet}
\bibinfo{author}{\bibfnamefont{{\O}.}~\bibnamefont{Elgar{\o}y}},
  \bibinfo{author}{\bibfnamefont{L.}~\bibnamefont{Engvik}},
  \bibinfo{author}{\bibfnamefont{M.}~\bibnamefont{Hjorth-Jensen}},
  \bibnamefont{and} \bibinfo{author}{\bibfnamefont{E.}~\bibnamefont{Osnes}},
  \bibinfo{journal}{Nucl. Phys. A} \textbf{\bibinfo{volume}{607}},
  \bibinfo{pages}{425} (\bibinfo{year}{1996}).

\bibitem[{\citenamefont{Baldo et~al.}(1998)\citenamefont{Baldo, Elgar{\o}y,
  Engvik, Hjorth-Jensen, and Schulze}}]{moderngap}
\bibinfo{author}{\bibfnamefont{M.}~\bibnamefont{Baldo}},
  \bibinfo{author}{\bibfnamefont{{\O}.}~\bibnamefont{Elgar{\o}y}},
  \bibinfo{author}{\bibfnamefont{L.}~\bibnamefont{Engvik}},
  \bibinfo{author}{\bibfnamefont{M.}~\bibnamefont{Hjorth-Jensen}},
  \bibnamefont{and} \bibinfo{author}{\bibfnamefont{H.-J.}
  \bibnamefont{Schulze}}, \bibinfo{journal}{Phys. Rev. C}
  \textbf{\bibinfo{volume}{58}}, \bibinfo{pages}{1921} (\bibinfo{year}{1998}).

\bibitem[{\citenamefont{Stairs et~al.}(2000)\citenamefont{Stairs, Lyne, and
  Shemar}}]{stairs}
\bibinfo{author}{\bibfnamefont{I.~H.} \bibnamefont{Stairs}},
  \bibinfo{author}{\bibfnamefont{A.~G.} \bibnamefont{Lyne}}, \bibnamefont{and}
  \bibinfo{author}{\bibfnamefont{S.~L.} \bibnamefont{Shemar}},
  \bibinfo{journal}{Nature} \textbf{\bibinfo{volume}{406}},
  \bibinfo{pages}{484} (\bibinfo{year}{2000}).

\bibitem[{\citenamefont{Cordes}(1993)}]{cordes93}
\bibinfo{author}{\bibfnamefont{J.}~\bibnamefont{Cordes}}, in
  \emph{\bibinfo{booktitle}{Planets Around Pulsars}}, edited by
  \bibinfo{editor}{\bibfnamefont{J.~A.} \bibnamefont{Phillips}},
  \bibinfo{editor}{\bibfnamefont{S.~E.} \bibnamefont{Thorsett}},
  \bibnamefont{and} \bibinfo{editor}{\bibfnamefont{S.~R.}
  \bibnamefont{Kulkarni}} (\bibinfo{publisher}{Astronomical Society of the
  Pacific}, \bibinfo{address}{San Francisco}, \bibinfo{year}{1993}), pp.
  \bibinfo{pages}{43--60}.

\bibitem[{\citenamefont{Shabanova et~al.}(2001)\citenamefont{Shabanova, Lyne,
  and Urama}}]{shabanova}
\bibinfo{author}{\bibfnamefont{T.~V.} \bibnamefont{Shabanova}},
  \bibinfo{author}{\bibfnamefont{A.~G.} \bibnamefont{Lyne}}, \bibnamefont{and}
  \bibinfo{author}{\bibfnamefont{U.~O.} \bibnamefont{Urama}},
  \bibinfo{journal}{Astrophys. J.} \textbf{\bibinfo{volume}{552}},
  \bibinfo{pages}{321} (\bibinfo{year}{2001}).

\bibitem[{\citenamefont{Link and Epstein}(2001)}]{linkepstein}
\bibinfo{author}{\bibfnamefont{B.}~\bibnamefont{Link}} \bibnamefont{and}
  \bibinfo{author}{\bibfnamefont{R.~I.} \bibnamefont{Epstein}},
  \bibinfo{journal}{Astrophys. J.} \textbf{\bibinfo{volume}{556}},
  \bibinfo{pages}{392} (\bibinfo{year}{2001}).

\bibitem[{\citenamefont{Jones and Andersson}(2001)}]{ja}
\bibinfo{author}{\bibfnamefont{D.~I.} \bibnamefont{Jones}} \bibnamefont{and}
  \bibinfo{author}{\bibfnamefont{N.}~\bibnamefont{Andersson}},
  \bibinfo{journal}{Mon. Not. Roy. Astr. Soc.} \textbf{\bibinfo{volume}{324}},
  \bibinfo{pages}{811} (\bibinfo{year}{2001}).

\bibitem[{\citenamefont{Rezania}(2003)}]{rezania1}
\bibinfo{author}{\bibfnamefont{V.}~\bibnamefont{Rezania}},
  \bibinfo{journal}{Astron. Astrophys.} \textbf{\bibinfo{volume}{399}},
  \bibinfo{pages}{653} (\bibinfo{year}{2003}).

\bibitem[{\citenamefont{Wasserman}(2003)}]{wasserman}
\bibinfo{author}{\bibfnamefont{I.}~\bibnamefont{Wasserman}},
  \bibinfo{journal}{Mon. Not. Roy. Astr. Soc.} \textbf{\bibinfo{volume}{341}},
  \bibinfo{pages}{1020} (\bibinfo{year}{2003}).

\bibitem[{\citenamefont{Cutler}(2002)}]{cutler}
\bibinfo{author}{\bibfnamefont{C.}~\bibnamefont{Cutler}},
  \bibinfo{journal}{Phys. Rev. D} \textbf{\bibinfo{volume}{66}},
  \bibinfo{pages}{084025} (\bibinfo{year}{2002}).

\bibitem[{\citenamefont{Cutler et~al.}(2003)\citenamefont{Cutler, Ushomirsky,
  and Link}}]{cul}
\bibinfo{author}{\bibfnamefont{C.}~\bibnamefont{Cutler}},
  \bibinfo{author}{\bibfnamefont{G.}~\bibnamefont{Ushomirsky}},
  \bibnamefont{and} \bibinfo{author}{\bibfnamefont{B.}~\bibnamefont{Link}},
  \bibinfo{journal}{Astrophys. J.} \textbf{\bibinfo{volume}{588}},
  \bibinfo{pages}{975} (\bibinfo{year}{2003}).

\bibitem[{\citenamefont{Ruderman et~al.}(1998)\citenamefont{Ruderman, Zhu, and
  Chen}}]{rzc}
\bibinfo{author}{\bibfnamefont{M.}~\bibnamefont{Ruderman}},
  \bibinfo{author}{\bibfnamefont{T.}~\bibnamefont{Zhu}}, \bibnamefont{and}
  \bibinfo{author}{\bibfnamefont{K.}~\bibnamefont{Chen}},
  \bibinfo{journal}{Astrophys. J.} \textbf{\bibinfo{volume}{492}},
  \bibinfo{pages}{267} (\bibinfo{year}{1998}).

\bibitem[{\citenamefont{Alpar et~al.}(1984)\citenamefont{Alpar, Langer, and
  Sauls}}]{als}
\bibinfo{author}{\bibfnamefont{M.~A.} \bibnamefont{Alpar}},
  \bibinfo{author}{\bibfnamefont{S.~A.} \bibnamefont{Langer}},
  \bibnamefont{and} \bibinfo{author}{\bibfnamefont{J.~A.} \bibnamefont{Sauls}},
  \bibinfo{journal}{Astrophys. J.} \textbf{\bibinfo{volume}{282}},
  \bibinfo{pages}{533} (\bibinfo{year}{1984}).

\bibitem[{\citenamefont{Sj{\"o}berg}(1976)}]{sjoberg}
\bibinfo{author}{\bibfnamefont{O.}~\bibnamefont{Sj{\"o}berg}},
  \bibinfo{journal}{Nucl. Phys. A} \textbf{\bibinfo{volume}{65}},
  \bibinfo{pages}{511} (\bibinfo{year}{1976}).

\bibitem[{\citenamefont{Chau et~al.}(1992)\citenamefont{Chau, Cheng, and
  Ding}}]{ccd}
\bibinfo{author}{\bibfnamefont{H.~F.} \bibnamefont{Chau}},
  \bibinfo{author}{\bibfnamefont{K.~S.} \bibnamefont{Cheng}}, \bibnamefont{and}
  \bibinfo{author}{\bibfnamefont{K.~Y.} \bibnamefont{Ding}},
  \bibinfo{journal}{Astrophys. J.} \textbf{\bibinfo{volume}{399}},
  \bibinfo{pages}{213} (\bibinfo{year}{1992}).

\bibitem[{\citenamefont{Mendell}(1998)}]{mendell}
\bibinfo{author}{\bibfnamefont{G.}~\bibnamefont{Mendell}},
  \bibinfo{journal}{Mon. Not. Roy. Astr. Soc.} \textbf{\bibinfo{volume}{296}},
  \bibinfo{pages}{903} (\bibinfo{year}{1998}).

\bibitem[{\citenamefont{Link and Cutler}(2002)}]{lc}
\bibinfo{author}{\bibfnamefont{B.}~\bibnamefont{Link}} \bibnamefont{and}
  \bibinfo{author}{\bibfnamefont{C.}~\bibnamefont{Cutler}},
  \bibinfo{journal}{Mon. Not. Roy. Astr. Soc.} \textbf{\bibinfo{volume}{336}},
  \bibinfo{pages}{211} (\bibinfo{year}{2002}).

\bibitem[{\citenamefont{Shaham}(1977)}]{shaham77}
\bibinfo{author}{\bibfnamefont{J.}~\bibnamefont{Shaham}},
  \bibinfo{journal}{Astrophys. J.} \textbf{\bibinfo{volume}{214}},
  \bibinfo{pages}{251} (\bibinfo{year}{1977}).

\bibitem[{\citenamefont{Epstein and Baym}(1992)}]{eb}
\bibinfo{author}{\bibfnamefont{R.~I.} \bibnamefont{Epstein}} \bibnamefont{and}
  \bibinfo{author}{\bibfnamefont{G.}~\bibnamefont{Baym}},
  \bibinfo{journal}{Astrophys. J.} \textbf{\bibinfo{volume}{387}},
  \bibinfo{pages}{276} (\bibinfo{year}{1992}).

\bibitem[{\citenamefont{Sonin}(1987)}]{sonin}
\bibinfo{author}{\bibfnamefont{E.~B.} \bibnamefont{Sonin}},
  \bibinfo{journal}{Rev. Mod. Phys.} \textbf{\bibinfo{volume}{59}},
  \bibinfo{pages}{87} (\bibinfo{year}{1987}).

\bibitem[{\citenamefont{Alcock et~al.}(1986)\citenamefont{Alcock, Farhi, and
  Olinto}}]{afo}
\bibinfo{author}{\bibfnamefont{C.}~\bibnamefont{Alcock}},
  \bibinfo{author}{\bibfnamefont{E.}~\bibnamefont{Farhi}}, \bibnamefont{and}
  \bibinfo{author}{\bibfnamefont{A.}~\bibnamefont{Olinto}},
  \bibinfo{journal}{Astrophys. J.} \textbf{\bibinfo{volume}{310}},
  \bibinfo{pages}{261} (\bibinfo{year}{1986}).

\bibitem[{\citenamefont{Link et~al.}(1999)\citenamefont{Link, Epstein, and
  Lattimer}}]{lel}
\bibinfo{author}{\bibfnamefont{B.}~\bibnamefont{Link}},
  \bibinfo{author}{\bibfnamefont{R.~I.} \bibnamefont{Epstein}},
  \bibnamefont{and} \bibinfo{author}{\bibfnamefont{J.~M.}
  \bibnamefont{Lattimer}}, \bibinfo{journal}{Phys. Rev. Lett.}
  \textbf{\bibinfo{volume}{83}}, \bibinfo{pages}{3362} (\bibinfo{year}{1999}).

\end{thebibliography}

\begin{figure}
\includegraphics{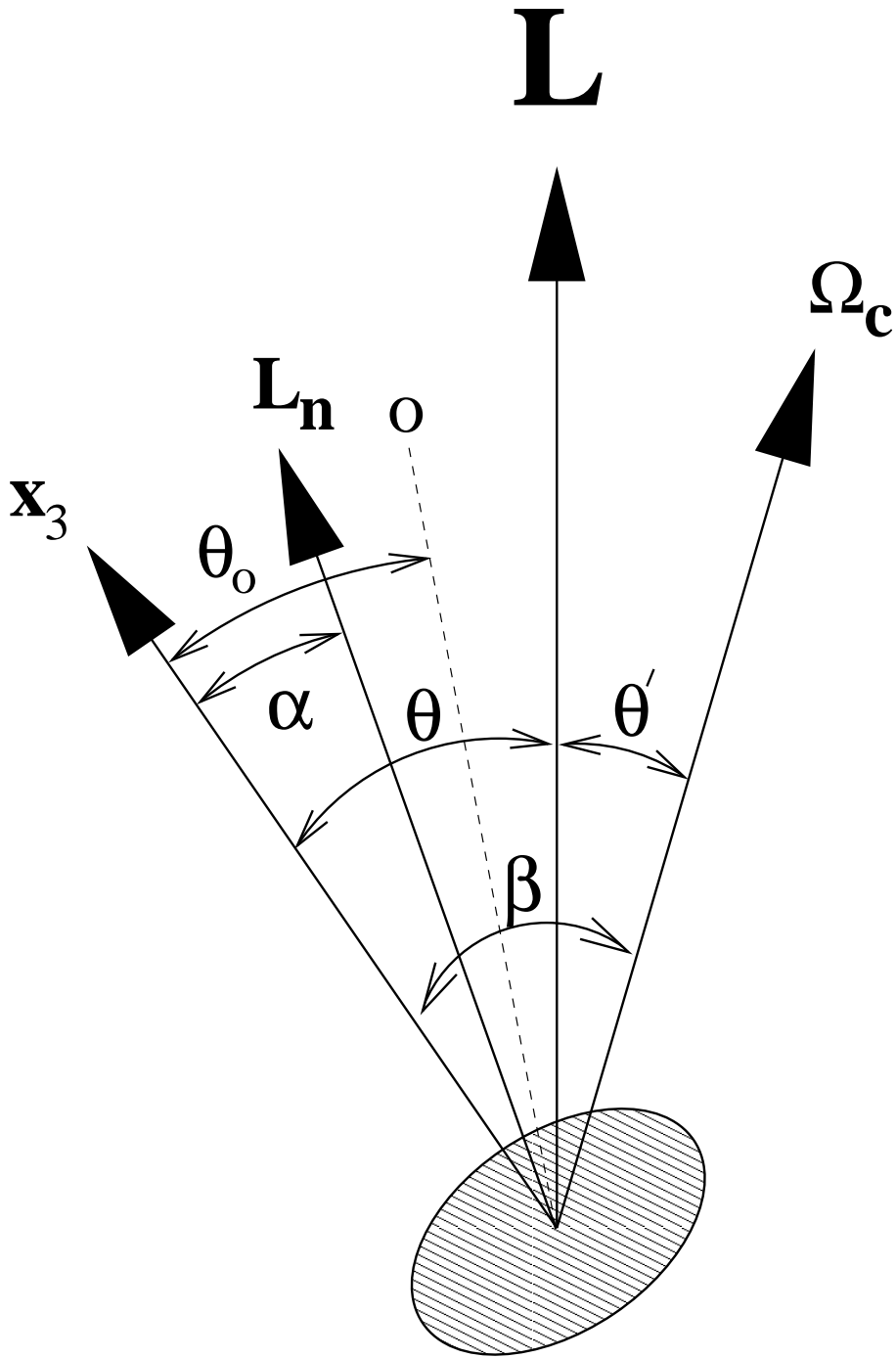}
\caption{\label{Fig1} The angles defined in the text. $\Omc$ takes a
circular path about axis $o$, the dashed line.}
\end{figure}

\end{document}